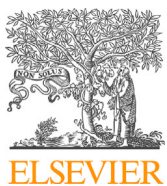

Contents lists available at ScienceDirect

# Optics and Lasers in Engineering

journal homepage: www.elsevier.com/locate/optlaseng

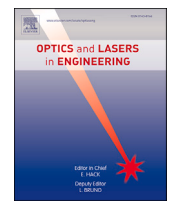

# Zonal shape reconstruction for Shack-Hartmann sensors and deflectometry

Jonquière Hugo *, Mugnier Laurent M., Michau Vincent, Mercier-Ythier Renaud

ARTICLE INFO

ABSTRACT

Some metrological means, such as Shack-Hartmann, deflectometry sensors or fringe projection profilometry, measure the shape of an optical surface indirectly from slope measurements. Zonal shape reconstruction, a method to reconstruct shape with a high number of degrees of freedom, is used for all of these applications. It has risen in interest with the use of deflectometers for the acquisition of high resolution slope data for optical manufacturing, especially because shape reconstruction is limiting in terms of shape estimation error.

Zonal reconstruction methods all rely on the choice of a data formation model, a basis on which the shape will be decomposed, and an estimator. In this paper, we first study the canonical Fried and Southwell models of the literature and analyze their limitations. We show that modeling the slope measurement by a point-wise derivative as they both do can induce a bias on the shape estimation, and that the bases on which the shape is decomposed are imposed because of this assumption.

In the second part of this paper, we propose to build an unbiased model of the data formation, without constraints on the choice of the decomposition basis. We then compare these models to the canonical models of Fried and Southwell.

Lastly, we perform a regularized MAP reconstruction, and compare the performance in terms of total shape error of this method to the state of the art for the Southwell and Fried models, first by simulation, then on experimental data. We demonstrate that the suggested method outperforms the canonical models in terms of total shape reconstruction error on a deflectometry measurement of the high-frequency content of a freeform mirror.

## 1. Introduction

Some metrological means, such as Shack-Hartmann, deflectometry sensors or fringe projection profilometry, measure the shape of an optical surface indirectly by means of slope measurements. In the context of optical manufacturing, the variable of interest for reworking an optical part is the shape map of the optical surface $\phi(x, y)$. Thus, the slope measurements acquired by a deflectometer or a Shack-Hartmann sensor are an indirect measure of the shape of the optical surface under study, and require the use of shape reconstruction algorithms. The study of the reconstruction problem represents an important issue for phase shift deflectometry because the shape reconstruction step in the data-processing chain is limiting in terms of shape reconstruction performance [1,2, Sect. 7].

Due to the use of Shack-Hartmann sensors for wavefront reconstruction [3], the problem of shape reconstruction from slope measurements has been a well-studied inverse problem since the 1980s. The first step to solve this problem is to choose a basis for the decomposition of the variable of interest. Usually, a distinction is made between modal meth-ods (decomposition on a basis of functions with extended support) and zonal methods (basis of functions with reduced support). Modal methods are characterized by a finite family of functions with extended support called modes, for example polynomial functions, splines, etc... These functional bases have been widely studied in the context of phase shift deflectometry [4–6]. While the state-of-the-art modal shape reconstruction methods have demonstrated good performance in the reconstruction of low and medium spatial frequencies, these bases limit the size of the solution space [4]. The low-pass filtering induced by this space restriction does not allow the exploitation of all the information measured by a deflectometric measurement, in particular the measurement of frequencies close to the sampling frequency, unless a complete basis is used, such as the Fourier basis. In contrast, in order to reconstruct the full set of spatial frequencies measured by phase-shift deflectometry or by a Shack-Hartmann, zonal methods use a shape decomposition basis composed of a large number of locally supported functions. While modal basis is popular in adaptive optics, as measuring the high spatial frequencies of the wavefront is sometimes not essential, some metrology applications, in particular optical manufacturing, require the high spatial frequencies to be estimated. Therefore, we will rule out modal bases in favor of zonal bases in this paper.

* Corresponding author.
*E-mail address:* hugo.jonquiere@gmail.com (H. Jonquière).

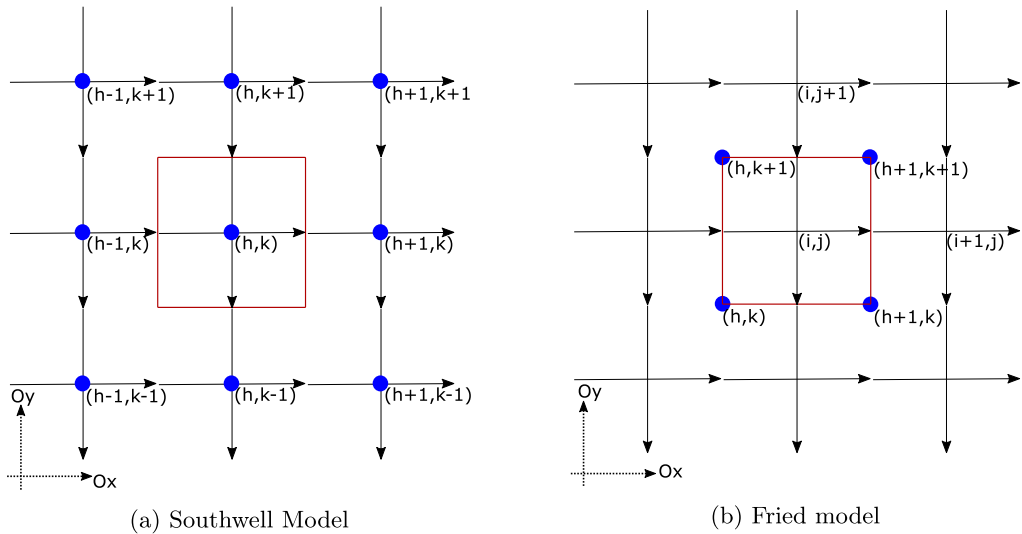

(a) Southwell Model   (b) Fried model

**Fig. 1.** Geometric configurations for a) the Southwell model and b) the Fried model.

The second step is to define a data formation model, named direct model or forward model to model the physical formation of the data. This model takes as an input the parameters of interest to generate the outcome of a measurement given these parameters. In the field of deflectometry, this issue has been completely neglected, both for zonal and modal bases. However, this issue is central because a data formation model that is far from the physics can induce systematic errors in the reconstructed shape. In this paper, we propose to design direct models corresponding to the physics of image formation, and to analyze the impact of the choice of the direct model on the total shape reconstruction error in an inverse problem formalism.

Finally, the third step corresponds to the choice of an estimator for solving the inverse problem corresponding to the direct model. The propagation of the slope measurement noise on the estimated shape is one of the major contributors to the total shape error [7]. The state of the art shape reconstruction algorithms in adaptive optics has very early used estimators based on regularization methods to limit the amplification of the measurement noise during the wavefront reconstruction [3, Chap. 5]. These estimators, such as Minimum Mean Square Error (MMSE) or Maximum A Posteriori (MAP) estimators, which are sometimes also used in the field of machine vision [8], have not, however, been adopted for shape reconstruction in deflectometry. In the last section of this paper, we propose the use of the so-called Maximum A Posteriori estimator, which, coupled with a direct model designed in this paper, leads to a systematic error-free shape reconstruction algorithm that is robust to the propagation of measurement noise on the estimated shape. We then experimentally validate the performance of this algorithm in terms of shape reconstruction error by comparison with experimental data acquired on a free-shape mirror by phase-shift interferometry.

## 2. Study of the canonical models for zonal shape reconstruction

Before diving into the discussion on the choice of a data formation model, we first discuss the preliminary question of the choice of the decomposition basis of the shape.

The problem of choosing a decomposition basis has been considerably studied in the context of deflectometry [9]. These studies, however, systematically neglect the question of the data formation model in favor of the question of the choice of the decomposition basis. Hence, all these algorithms are based on the same data formation model [9,1,7], the Southwell model described in the section below. Because an erroneous data formation model leads to systematic errors on the reconstructed form, we propose in this section to study the canonical models of data formation, in order to then choose a shape reconstruction algorithm which minimizes the total shape reconstruction error.

Three canonical slope formation models have been widely studied: the Southwell model, the Fried model, and the Hudgin model [10–12]. These models differ in their geometrical configurations and in the chosen decomposition basis. The Hudgin configuration corresponds to a model where the two components of the slope vector field **s** are measured at two distinct positions [12]. In the case of an experimental deflectometric measurement or a Shack-Hartmaan sensor, the two components of the slope vector field of the optical surface **s** are measured at the same position. This model does not correspond to the physics of the data acquisition, so the use of the Hudgin model will generate a systematic error on the reconstructed shape. For this reason, we will study in this paper only the Southwell and Fried configurations. Fig. 1 represents the geometric configurations associated with the Southwell and Fried models. The intersection of the arrows indicates the sampling point of the two slope components. The blue dot indicates the sampling of the points of the desired shape.

We emphasize that these models assume a pointwise measurement of slopes, hence the data corresponds to the pointwise analytical derivative of the sought shape. In order to properly differentiate the models developed in this paper from the canonical Southwell and Fried models, the latter will be referred to in this manuscript as the Pointwise Southwell model (PS) and the Pointwise Fried model (PF).

In the Southwell geometric configuration, the shape sampling grid points are located at the same positions as the slope measurement locations, while in the Fried geometric configuration, the shape sampling grid is shifted by half a pixel along the axes relative to the slope sampling. In this sense, Fried's model assumes that the shape is sampled on the corners of the slope detector pixels. We can emphasize that the notion of pixel is used here to describe the position of the different samplings but these models being pointwise, they do not involve a detector pixel per se. In addition to the geometrical configurations, these two direct models also differ by their choice of the basis of the shape decomposition.

### 2.1. Pointwise Southwell model

The pointwise Southwell (PS) model is based on a decomposition of the shape into a quadratic spline [11], which assumes that the shape of the optical surface on the segment connecting two slope sampling points is a polynomial of order 2. Because the shape is decomposed on a support consisting of separated segments, it is locally defined by a polynomial defined on a support of vertical and horizontal lines. The PS model does not have an explicit continuous shape in 2-dimensions, and the user needs to define continuous two dimensional shape if needed.

Thus, we obtain for example on the Ox axis:

$$\forall x \in [0, p], \phi(x) = ax^2 + bx + c,$$

where $a, b$ and $c$ are polynomial coefficients and $p$ the sampling step of the slope. This model is pointwise, *i.e.* it assumes that the measured slope corresponds to the sampling at a point of a continuous gradient of the shape of the optical surface, and that the quantity of interest, the shape itself, also corresponds to the sampling at a point of a continuous function. By numbering $(h, k)$ the point of the sampling grid located in $(0,0)$, we obtain:

$$\begin{cases} s_x(h, k) = b \\ s_x(h+1, k) = 2ap + b \\ \phi(h, k) = c \\ \phi(h+1, k) = ap^2 + bp + c \end{cases},$$

where $s_x$ is the component along the Ox axis of the slope field **s**. This leads to the well-known equation describing the formation of the data along the Ox axis in the Southwell model:

$$\frac{s_x(h+1, k) + s_x(h, k)}{2} = \frac{\phi(h+1, k) - \phi(h, k)}{p}. \tag{1}$$

By an analogous reasoning along Oy, we obtain:

$$\frac{s_y(h, k+1) + s_y(h, k)}{2} = \frac{\phi(h, k+1) - \phi(h, k)}{p}. \tag{2}$$

These equations derived by Southwell in his seminal paper in 1980 [11] and subsequently used in solving the inverse problem raise several points. On the one hand, the equations describing the formation of the data do not directly calculate the value of the slope at a point for a given shape, but generate a system of coupled linear equations. It is therefore not a classical direct model of the type $\mathbf{s} = \mathcal{F}(\phi)$ (see [Chap. 1][13]).

It can be seen from equations (1) and (2) that this model can, on the other hand, be considered a direct model with respect to an effective measurement of the pre-processed slope data **s'**:

$$\mathbf{s'} = \mathbf{A}_{SP}\mathbf{s},$$

where $\mathbf{A}_{SP}$ the matrix performing the local averaging of the slopes described in Equations (1) and (2). The pre-processing matrix thus creates an average slope point shifted by half a pixel, located between the two original slope points. This averaging is performed along two different directions for the two measured slope components, thus this pre-processing is an-isotropic. Finally, this model makes the assumption of a shape decomposable on a basis of quadratic 1-d spline functions. This assumption generates systematic errors if the residual of the projection of the optical surface shape on this basis is non negligible.

### 2.2. Pointwise Fried model

The pointwise Fried model (PF) is based on a decomposition of the shape into bi-linear splines [10,11,14], which assumes that the shape of the optical surface within a square composed of four slopes is:

$$\forall (x, y) \in [0, p]^2 \; \phi(x, y) = ax + by + cxy + d, \tag{3}$$

where $a, b, c$ and $d$ are polynomial coefficients and $p$ the sampling step of the slope. Like Southwell's model, this model is pointwise, *i.e.* it assumes that the measured slope corresponds to the sampling at a point of a continuous gradient. Indeed, identifying by $(i, j)$ the slope sampling grid point located in $(i, j) = (0, 0)$, and by $(h, k)$ the shape sampling grid point located in $(-\frac{p}{2}, -\frac{p}{2})$, we obtain:

$$\begin{cases} s_x(i, j) = a + c\frac{p}{2} \\ \phi(h+1, k) - \phi(h, k) = ap \\ \phi(h, k+1) - \phi(h, k) = bp \\ \phi(h+1, k+1) - \phi(h, k+1) = ap + cp^2 \end{cases}.$$

This leads to the equation describing the formation of data along Ox in the Fried model:

$$s_x^{i,j} = \frac{1}{2p} \left( \phi_{h+1,k} - \phi_{h,k} + \phi_{h+1,k+1} - \phi_{h,k+1} \right). \tag{4}$$

And by analogous reasoning along Oy:

$$s_y^{i,j} = \frac{1}{2p} \left( \phi_{h+1,k} - \phi_{h,k} + \phi_{h+1,k+1} - \phi_{h,k+1} \right). \tag{5}$$

Unlike the PS model, the PF model leads to an equation that allows the direct calculation of measured slopes from any shape. Moreover, the definition support of the polynomials related to the decomposition into bi-linear splines is two-dimensional and identical for both components. This model assumes a shape decomposable on bi-linear splines, and thus a global shape of class $C^0$. Depending on the residual error of the projection of the optical surface shape on this basis, this assumption can generate systematic errors.

### 2.3. Remarks and discussion on the definition of pointwise Fried's and Southwell's models

The choice made by the authors to use bi-linear forms for the Fried model or quadratic forms for the Southwell model is dictated here by the number of degrees of freedom that these models can support. The choice of a linear spline for the Southwell model would imply that the slopes $s_x(i, j)$ and $s_x(i+1, j)$ are equal. In contrast, choosing a bi-quadratic spline shape decomposition for the PF model would imply that the number of polynomial coefficients to be estimated on this support exceeds the number of shape points adjacent to a slope point. Thus, through the number of degrees of freedom, *i.e.* through the number of shape points influencing the value of a slope point in these models, the choice of a geometry constrains the choice of a decomposition into basis functions. For example, the approach of Huang et al. [15] using a PS data formation model and a base of 4th order spline functions is constrained to use the value of the 4 shape points neighboring the considered slope point, still in a one-dimensional model.

This constraint on the choice of basis functions brought by the definition of the data formation models is problematic. Indeed, we would like to have a real choice of the shape decomposition basis, for example to model the frequency aliasing phenomenon, or to choose locally supported bases of class $C^\infty$.

To our knowledge, no algorithm in the literature for zonal reconstruction in deflectometry uses the Fried [9,1,7] model. The Southwell model is valued for the correspondence between the slope and shape [7] sampling grids, and the idea that a shape decomposition on a quadratic spline basis would be more optimal than a shape decomposition on a bi-linear spline basis. Contrary to Huang et al. [7] claim's, it is not clear that using a quadratic spline model guarantees Southwell's model a lower systematics error. Indeed, and as Southwell already pointed out [11], the problem of interpolating a continuous shape should not be confused with the problem of reconstructing a grid of shape points from discrete measurements.

While the bibliography on basis function selection is extensive (see e.g. Section 2 of this paper), the model for data formation in deflectometry is consistently the SP [9,1,7] model. Yet, both the PS and PF models make the assumption of pointwise slope measurements, which does not correspond to the physics of data formation. Indeed, the acquired slope data physically correspond to the average of the slope value on a sub-aperture for a Shack-Hartmann sensor, or on a detector pixel for a deflectometer. Moreover our analysis shows that the PS model is not a direct model in the strict sense of the word, as it relies on an overlooked anisotropic preprocessing that filters the slope data.

## 3. Building an unbiased & non-limiting model of data formation

In this section, we propose to design direct models corresponding to the physics of data formation, and to analyze the impact of the choice of the direct model on the total error of shape reconstruction in an inverse problem formalism. To our knowledge, this approach is original in the field of zonal reconstruction in deflectometry and adaptive optics.

### 3.1. Physics-based model of the measured slope

Consider the value of the slope associated with a detector pixel in the case of deflectometry or a sub-aperture of a Shack-Hartmann sensor, corresponding to one element of a discrete set of slopes measured from a continuous shape (or phase). The physical process of forming the slope data corresponds to averaging the value of the gradient of the shape to be reconstructed on a surface that we will call the integration support [3, Chap. 5]. In the case of deflectometry, this surface corresponds to the surface of the optical part optically conjugated to the surface of a detector pixel, while in the case of a Shack-Hartmann sensor, it corresponds to the surface of the sub-aperture of the Shack-Hartmann sensor. Therefore, the horizontal component of the slope $s_x$ associated with the integration surface (or pixel) $(h, k)$ is [3]:

$$s_x^{h,k} = \frac{1}{p^2} \int_{-\frac{p}{2}}^{\frac{p}{2}} \int_{-\frac{p}{2}}^{\frac{p}{2}} \frac{\partial \phi}{\partial x}(x + hp, y + kp) \, dx dy, \tag{6}$$

where $p$ is the width of a pixel, $\phi$ is the continuous shape of the optical surface under inspection, and the origin of the reference frame $O_{xy}$ is taken at the center of the pixel $(0,0)$. Following, we will refer to this model as a model with integral approach. Since the shape is continuous, we obtain the model [14]:

$$s_x^{h,k} = \frac{1}{p^2} \int_{-\frac{p}{2}}^{\frac{p}{2}} \left[ \phi\left((h + \frac{1}{2})p, y + kp\right) - \phi\left((h - \frac{1}{2})p, y + kp\right) \right] dy. \tag{7}$$

It can be seen that under the assumption of a continuous shape, and independently of the choice of the decomposition basis, the local slope measured along Ox corresponds to the difference of the integral of the shape along the two vertical boundaries of the integration support. By symmetrical reasoning, it can be shown that the local slope measured along Oy corresponds to the difference of the shape integral along the two horizontal boundaries of the integration support. The dependence of direct models on the values of the basis functions on the boundaries of the integration support demonstrated in Equation (7) implies, in particular, that the sampling geometry of the slope relative to the integration support, i.e., the choice of a Southwell or Fried type geometric configuration, is of particular importance for these direct models. Hence, we define two integral approach models, respectively Integral Approach Southwell (IAS) model and Integral Approach Fried (IAF) model.

Fig. 2 shows the Southwell and Fried geometric configurations for the IAS & IAF models. The intersection of the arrows indicates the center of the slope measurement (center of the detector pixel). The blue dots indicate the centers of the local basis function of the shape decomposition. The red squares indicate the boundaries of the detector pixels. The green circle represents an example of a finite support chosen for the local basis functions of the shape decomposition. Recall that the area of the latter was assumed to be at least the size of the area of the latter, but not more than four times the area of the latter. Note that although in Eq. (6) & (7) the pixels of the detector are assumed to be a square array, the IAF & IAS models are not limited to square geometries, and can be extended to any given geometrical configuration. In practice, due to alignment or aberration in the optical system, the integration support of the slopes measurement can be distorted (e.g., quadrilateral). For these cases, the following calculations need to be updated for the corresponding grid geometry. Note that for a rectangular integration support, the model correction is straightforward, and simply consists in considering different pixel widths $p_x$ and $p_y$ for the two axes $O_x$ and $O_y$. The origin of the datum $O_{xy}$ is taken at the center of the slope pixel for both configurations ($(h, k)$ for the Integral Approach Southwell model, and $(i, j)$ for the Fried Integral Approach model).

In the IAS model, the local functions of the shape decomposition basis are centered on the center of the detector pixel, while in the IAF model, these functions are centered on the corners of the integration medium.

In the following section, we will calculate the discrete model we obtain by choosing a basis of decomposition for the shape.

### 3.2. Decomposition of the shape on a basis of local support functions

As explained in the introduction to this section, a direct model is characterized by a number of assumptions about the physics of data formation and a basis for shape decomposition. Because we are looking for a local shape decomposition that allows us to exploit the slope information over the set of measured spatial frequencies, we wish to decompose the shape on a basis of locally supported continuous functions. The supports of the basis functions related to a zonal model must partially overlap to reconstruct a continuous shape without systematic error, but in the idea of preserving a basis of local functions, we make the assumption in this calculation that the shape is decomposed on a set of identical functions $\phi_0$ translated and centered on pixels $(i, j)$, which we note $\phi_{i,j}(x, y) = \phi_0(x - ip, y - jp)$:

$$\phi(x, y) = \sum_{i=1}^{N'} \sum_{j=1}^{M'} c_{i,j} \phi_{i,j}(x, y) = \sum_{i=1}^{N'} \sum_{j=1}^{M'} c_{i,j} \phi_0(x - ip, y - jp), \tag{8}$$

where $N'M'$ the number of points of the decomposition of the shape $\phi$. In order to minimize the dependencies in the direct model, we assume that the width of the local support of the function $\phi_0$ is between one and two times the width of the integration support. We also assume, per simplicity, that this function is symmetric with respect to the axes $O_x$ and $O_y$. Finally, without loss of generality, we impose a normalization condition on the decomposition basis, i.e. that $\phi_0(0, 0) = 1$. Thus, taking into account the assumption on the width of the local support, we have:

$$\phi(hp, kp) = c_{h,k} \tag{9}$$

Note that to avoid discontinuity errors in the shape $\phi$ estimation, the function $\phi_0$ must be continuous and must cancel at the boundary of its support.

#### 3.2.1. Equation of the IAF direct model

The calculation being too long, it is given in Appendix A, and we obtain a model:

$$s_x^{i,j} = \alpha(\frac{c_{i+1,j} - c_{i,j} + c_{i+1,j+1} - c_{i,j+1}}{2p}). \tag{10}$$

We recognize an equation proportional to the equation derived from the Fried model, within a coefficient $\alpha$ which depends only on the choice of the basis of decomposition of the shape. Additionally, we demonstrate the point-wise Fried model is a specific case of the IAF model in Appendix C.1. The Point-wise Fried model is thus in good agreement with the formation of the data in a deflectometry setup, and thus does not generate a systematic modeling error in the shape reconstruction problem.

#### 3.2.2. Equation of the IAS direct model

Calculation is given in Appendix B as well, and we obtain:

$$s_x^{i,j} = \beta(\frac{c_{i+1,j} - c_{i-1,j}}{2p}). \tag{11}$$

The IAS model results in a different equation than the Southwell model, which does not incorporate pre-processing of the slope data, and thus does not filter the data, unlike the PS model. Additionally, unlike the Southwell model, which relied on an-isotropic pre-processing of the slopes, the IAS model is symmetric. In Appendix D, we demonstrate that the point-wise Southwell model is not a specific case of the IAS model. Hence, the PS model generates systematic modeling error in the estimated shape.

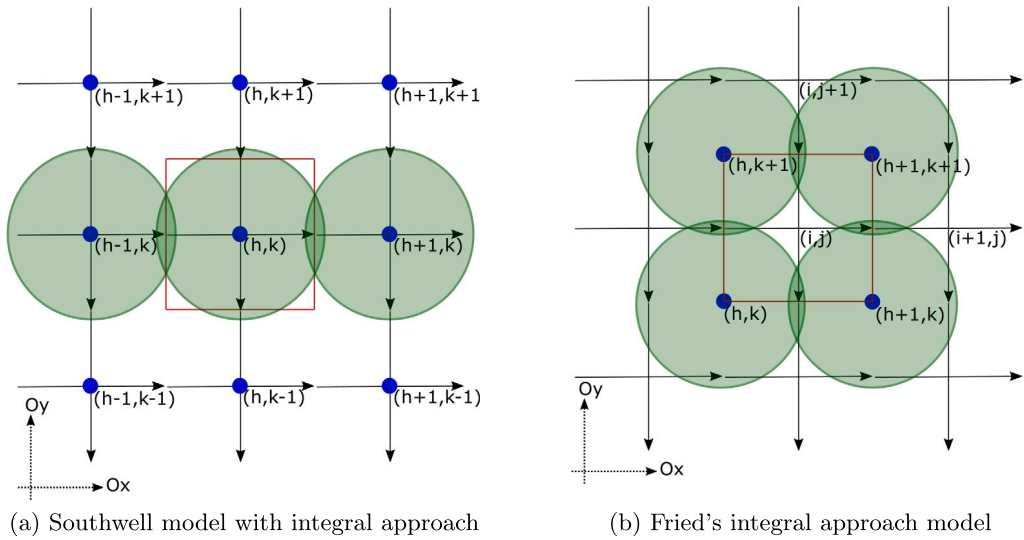

(a) Southwell model with integral approach
(b) Fried's integral approach model

**Fig. 2.** Geometric configurations for (a) the Integral Approach Southwell model and (b) the Integral Approach Fried model. The origin of the datum $O_{xy}$ is taken at the center of the slope pixel for both configurations (($h, k$) for the Southwell Integral Approach model, and ($i, j$) for the Fried Integral Approach model).

### 3.3. Note on the choice of $\phi_0$

The IAF and IAS models both depend on a constant (respectively $\alpha$ and $\beta$) which depends only on the choice of the basis function $\phi_0$. This coefficient characterizes the equation associated with the IAS and IAF models, and the question of the choice of this coefficient arises.

In this paper, we have dealt with the bases of continuous shape decomposition without ever worrying about the surjectivity of the bases used. In order for our physical model to correctly describe the formation of the data, it is necessary that the continuous shape we are looking for belong to the vector space of functions generated by the locally supported function base.

Note that a tilt shape (a polynomial of order 1) belongs to this vector space only if $\alpha = 1$ in the IAF case and $\beta = 1$ in the IAS case. Hence, for basis that can perfectly fit a tip or a tilt:

$$\begin{cases} \alpha = 1 \\ \beta = 1. \end{cases} \tag{12}$$

#### 3.3.1. Conclusion: direct models & integral approach

We have shown in the previous section that the PS model is not strictly speaking a direct model. Indeed, the underlying assumptions of the PS data formation model induce an anisotropic pre-processing of the measured slope data. Even though this pre-processing filters out part of the measurement noise, it also alters the high spatial frequencies of the slope data, which induces a systematic error, *i.e.* a bias on the reconstructed shape as will be shown in the simulations described in Section 4.

Furthermore, the FP and PS models are both non-physical data formation models, as they assume that the slope measurement is pointwise. To avoid systematic errors, we designed two original models, respectively IAF and IAS, based on an integral approach, and thus in agreement with the physics of data formation. These direct models impose minimal constraints on the choice of basis functions, and allow one to study the impact of the assumptions underlying the FP and PS models in a common formalism.

We have shown that the IAF model leads to a discrete equation proportional to that of the PF model. This PF model can therefore be interpreted as a particular IAF model, and is therefore in agreement with the physics of data formation. On the contrary, the IAS model leads to a discrete equation that is incompatible with the PS model. The PS model therefore generates a systematic error (bias) on the reconstructed shape. The IAS model leads to a symmetric discrete equation, but limits the choice of basis functions to a subset with restricted support (as demonstrated in Appendix B).

From the point of view of the definition of the direct model in the context of zonal shape reconstruction, we do not recommend the use of the PS model, although it is currently the most widely used model in the deflectometry literature, and prefer for all the reasons mentioned above the proposed IAS and IAF models. Simulations in later Section 4 of this paper demonstrate confirm this result.

## 4. Investigating the performance of these models by means of simulation

Once a direct model is defined, it is necessary to choose an estimator to solve the associated inverse problem. A zonal shape reconstruction algorithm corresponds to the numerical implementation of an estimator associated to a direct problem. This section explains the problems related to the choice of an estimator associated to a direct model (PS, PF, IAS or IAF), in particular the propagation of the measurement noise through the shape reconstruction. We recall that the criteria for choosing the estimator and the direct model for zonal shape reconstruction are the propagation of noise and the systematic error (or bias) of shape reconstruction. These two criteria make up the total shape reconstruction error as the Mean Square Error is the sum of the variance and of the squared bias.

### 4.1. Least square estimator

The most popular estimator in deflectometry is the well-known least squares estimator [15,7,16].

#### 4.1.1. Application of the least squares estimator to shape reconstruction

Since the equations describing the direct models PS, PF, IAS and IAF are linear (see Equations (1), (4), (10) and (11)), it is possible to write them in matrix shape. Respectively, for the PS model [11,10]:

$$\mathbf{A}_{SP}\mathbf{s} = \mathbf{D}_{SP}\boldsymbol{\phi} + \mathbf{A}_{SP}\mathbf{n}, \tag{13}$$

and for the PF, IAF and IAS models (see Equations (4), (10), (11) and (10)):

$$\mathbf{s} = \mathbf{D}_m \boldsymbol{\phi} + \mathbf{n}, \qquad (14)$$

where $\mathbf{s} = (s_x, s_y)$ is the vector of the measured slope fields of the optical surface, $\boldsymbol{\phi}$ is the shape of the optical surface, $\mathbf{D}_m, m \in [\text{PF, IAF, IAS}]$ and $\mathbf{D}_{SP}$ matrices represent the linear model of data formation, $\mathbf{A}_{SP}$ is the matrix averaging the adjacent slopes in Equation (13) and $\mathbf{n}$ is a realization of the unavoidable noise. The least squares estimator associated with the PS model is then:

$$\hat{\boldsymbol{\phi}} = \arg\min_{\boldsymbol{\phi}} \left( \|\mathbf{A}_{SP}\mathbf{s} - \mathbf{D}_{SP}\boldsymbol{\phi}\|^2 \right), \qquad (15)$$

and that associated with the PF, IAF and IAS models:

$$\hat{\boldsymbol{\phi}} = \arg\min_{\boldsymbol{\phi}} \left( \|\mathbf{s} - \mathbf{D}_m\boldsymbol{\phi}\|^2 \right). \qquad (16)$$

### 4.1.2. Errors associated with the use of a least squares estimator for the PS model

We are interested here in the case of the least squares estimator associated with the PS model. However, we have seen in the Section 2 that the PS model induces an an-isotropic pre-processing of the slopes which gives rise to pseudo-measurements of slopes. This pre-processing acts as a filtering of the slope measurement noise, but also of the slope data, in particular of the high spatial frequencies. While this filtering improves the propagation of noise, it also biases the PS model by filtering out some of the measured information, and studying the performance of the PS model without taking into account the total reconstruction error is therefore a methodological error. To our knowledge, the performance of the Southwell model from the point of view of noise propagation is not identified in the literature as corresponding to a trade-off between the bias and the variance, and the error on the reconstructed shape associated with the PS model has in fact been systematically underestimated in the literature. Thus, Zou et al. recommend using the Southwell model rather than the Fried model from the point of view of measurement noise propagation by studying only the variance of the noise on the reconstructed shape [17]. Similarly, C. Correia has studied in detail the different zonal shape reconstruction methods and points out the good performance of the Southwell model with respect to noise propagation without mentioning the underlying filtering of the slope data [14, Sect. 4].

Moreover, for a white, isotropic and Gaussian noise on the slopes, the effective noise on the pseudo-slopes $\mathbf{A}_{SP}\mathbf{n}$ is no longer white: its Power Spectral Density (PSD) is proportional to the squared transfer function of the operator $\mathbf{A}_{SP}$. The least squares estimator then no longer corresponds to the maximum likelihood [18]. The bibliography concerning deflectometry makes to our knowledge no mention of maximum likelihood estimator, and the methods for solving the PS model are based on a least squares estimator [15,7,16]. It is necessary when using the PS model to study the performance of this model from the point of view of the total error on the reconstructed shape, *i.e.* by including both variance and bias in the error calculation.

### 4.2. Maximum A Posteriori estimator

Zonal least squares reconstruction is a well known ill-conditioned problem [8,3], in which noise propagation has been identified as a phenomenon limiting the high frequency reconstruction performance [14, 2]. The bibliography associated with adaptive optics suggests to use Maximum A Posteriori (MAP) estimators for shape reconstruction, by modal or zonal methods. Indeed, if the prior and noise distributions are both Gaussian distributions, the MAP estimator is identical to the Minimum Mean-Square Error estimator [19, Sect. 2.4]. The Maximum A Posteriori (MAP) estimator to regularize the shape reconstruction is expressed as:

$$\hat{\boldsymbol{\phi}} = \arg\min \left( \frac{1}{2\sigma^2} \|\mathbf{D}\boldsymbol{\phi} - \mathbf{s}\|^2 + \frac{1}{2} R(\phi) \right), \qquad (17)$$

where $\mathbf{D}$ is the matrix associated with the chosen direct model, and $R(\phi)$ is a regularization term. Adopting a centered Gaussian *a priori* probability distribution for the shape leads to the following regularization term:

$$R(\phi) = \lambda \sum_{k_x} \sum_{k_y} \frac{|\tilde{\phi}(k_x, k_y)|^2}{S_\phi(k_x, k_y)}, \qquad (18)$$

where $\mathbf{k} = (k_x, k_y)$ is the spatial frequency vector, $\lambda$ a regularization coefficient weighting the influence of the regularization term with respect to the least squares term, $\tilde{X}$ denotes the Fourier transform of a variable $X$ and $S_\phi(\mathbf{k})$ the *prior* shape Power Spectral Density (PSD). We assume, following [20], that the shape PSD follows for optical manufacturing applications an inverse power law:

$$S_\phi(\mathbf{k}) \propto \frac{1}{\|\mathbf{k}\|^2}. \qquad (19)$$

The regularization term $R(\phi)$ is penalizing the ratio of the squared modulus of the Fourier transform of the shape to a PSD. This regularization term is expressed in the Fourier domain, but as we are using a 2-power law, we can express this regularization term in the direct domain using the Parseval theorem [21]:

$$R(\phi) \propto \lambda \|\nabla\phi\|^2, \qquad (20)$$

where $\nabla$ denotes the discrete gradient operator. We underline that expressing the regularization term in the direct domain allows for shape reconstruction on any given aperture using a sparse matrix formalism [13]. The regularization coefficient $\lambda$ which minimizes the mean-square error can be derived analytically from the prior on the power spectral density of the shape to be reconstructed [21,13]. This result induces that no trial-and-error is needed with regards to the $\lambda$ coefficient.

### 4.3. Simulation of the performance in terms of mean square error on the reconstructed shape

In order to assess the performance a direct model/estimator pair, it is necessary to perform numerical simulations quantifying the total shape error due to the sum of the systematic shape reconstruction error (or bias) and the propagation of measurement noise for each of these model/estimator pairs. Assuming periodic boundary conditions, and using a convolution formalism, we analytically derive the total quadratic shape error (bias & variance) of a MAP estimator averaged on a large number of noise and shape realizations given a prior on the power spectral density (PSD) of shape to be reconstructed and the noise distribution standard deviation on the measurement. As the calculation is long, it is postponed in Appendix E, and we obtain:

$$\epsilon_\phi^2 = \frac{1}{N^2} \sum_{m_x=1}^{N} \sum_{m_y=1}^{N} \left( \frac{\lambda \frac{\sigma^2}{S_\phi(k_x, k_y)}}{\left|\tilde{h_x}\right|^2 + \left|\tilde{h_y}\right|^2 + \lambda \frac{\sigma^2}{S_\phi(k_x, k_y)}} \right)^2_{k_x, k_y} S_\phi(k_x, k_y)$$

$$+ \frac{\sigma^2}{N^2} \sum_{m_x=1}^{N} \sum_{m_y=1}^{N} \left( \frac{\left|\tilde{h_x}\right|^2 + \left|\tilde{h_y}\right|^2}{\left(\left|\tilde{h_x}\right|^2 + \left|\tilde{h_y}\right|^2 + \lambda \frac{\sigma^2}{S_\phi(k_x, k_y)}\right)^2} \right)_{k_x, k_y},$$

where $(h_x, h_y)$ are the convolution kernel specific to the model considered and $\tilde{X} = \mathfrak{F}(x)$ denotes the Fourier transform. This analytical development of the MAP total quadratic shape error displays a clear distinction between a systematic error term (left) and a noise propagation term (right). For $\lambda \to 0$, the MAP estimator converges towards the least squares estimator and is dominated by the noise propagation term. For $\lambda \to +\infty$, the total quadratic error is dominated by the *a priori* power spectral density of the shape. Optimal regularization coefficient can be demonstrated to be 1 if the *a priori* distribution is correctly chosen [13, Chap. 4]. Please note that due to the pre-processing, the PS model leads to a slightly different total shape error equation, displaying a systematic error even for $\lambda = 0$. The power spectral density of the shape is chosen such as:

$$\sigma_\phi = \|\phi\| = 200 \text{ nm},\tag{21}$$

which corresponds to the typical RMS shape error of an optical surface in the deflectometry field.

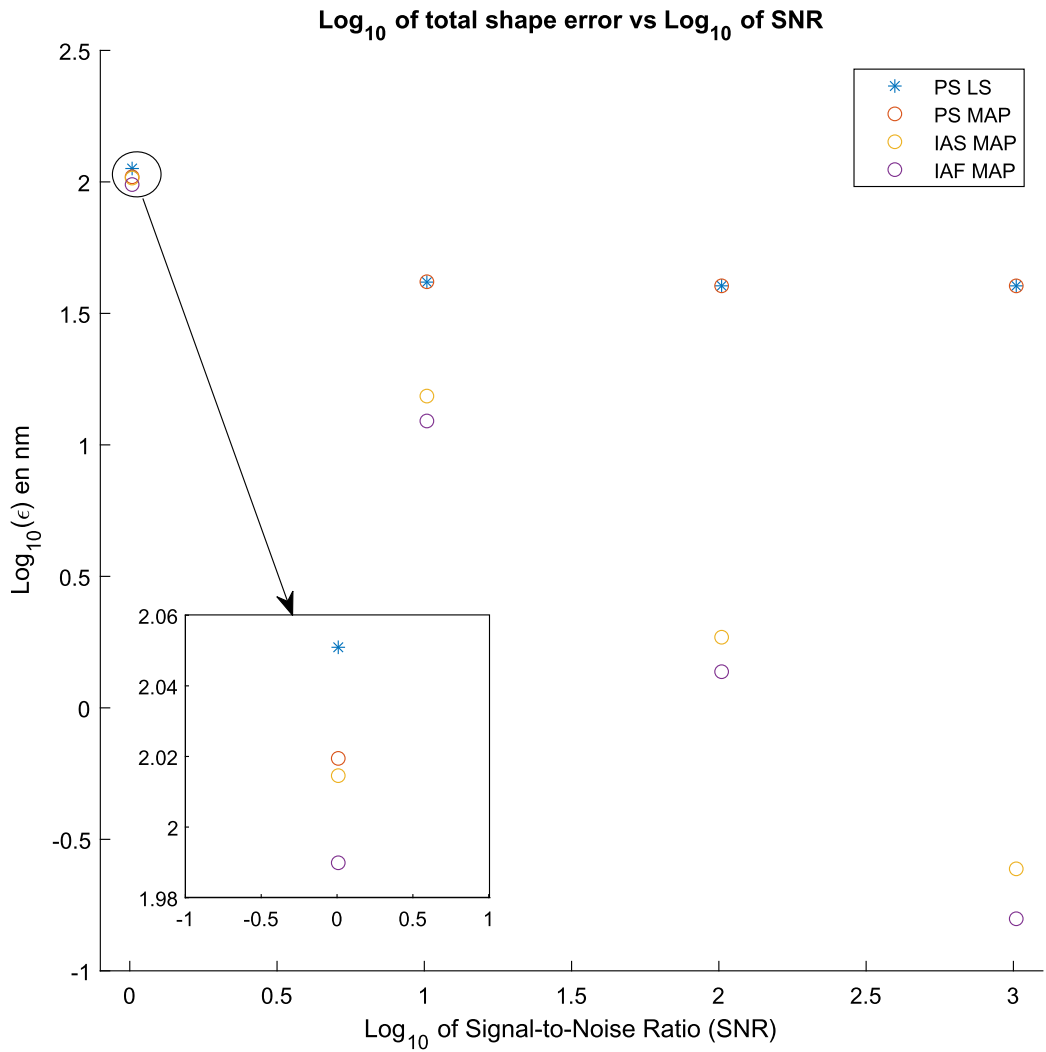

**Fig. 3.** Simulation of the total error on the reconstructed shape as a function of the signal-to-noise ratio (SNR) for the pairs direct model & PS LC, PS MAP, IAS MAP, and IAF MAP estimator and for different Signal-to-Noise-Ratio (SNR) values. Root-mean-square of the shape to be reconstructed is 200 nm. We underline that the PF model is identical to the IAF model for well-chosen basis.

We therefore performed a series of numerical simulations and computed the total error on a reconstructed shape for different values of Signal-to-Noise-Ratio (SNR) for the various direct model/estimator pairs. Following the formalism defined in the section 4.2, we assume a PSD of the estimated form defined by Equations (21) and (19). From the PSD of the shape, we can calculate the PSD of the slopes given a forward model. Using the Parseval theorem, we then define the signal-to-noise ratio SNR of the data as the ratio between the norm of the slopes and the standard deviation of the measurement noise on the slopes $\sigma$, assumed to follow a known, white, Gaussian distribution:

$$\text{SNR} = \frac{\|s\|}{\sigma}.\tag{22}$$

Because the PSD of the slope is fixed by the choice of the prior PSD of the shape, imposing the value of the SNR therefore set up the noise value $\sigma$. The contribution of the mean value has been subtracted from the set of estimated mean square errors because this mode is unseen by the set of direct models considered. Following [21] & [13], the regularization coefficient is then calculated to minimize the mean-square error. The sensitivity to the value of this regularization coefficient is studied in the reference [13, Chap. 4]: this study demonstrates the slow, weak, typically logarithmic, dependency of the MSE with $\lambda$.

Fig. 3 represents the total error on the reconstructed shape computed by numerical simulation as a function of the signal-to-noise ratio (SNR) for the pairs of direct model & estimator PS LS, PS MAP, IAS MAP, and IAF MAP and for different values of SNR. The total error, or mean-square-error (MSE) on the estimated shape is the quadratic sum of the bias and of the noise contaminating the measure propagated through the reconstruction.

The simulations presented in Fig. 3 show that the IAF/MAP model estimator pair performs better in terms of MSE than all the other estimator/model pairs for any given SNR. The MSE associated with the IAF model is lower than the MSE associated with the IAS model for any given SNR. The observed MSE in this case is correlated to the number of unseen modes by the direct model considered. The more the number of unseen modes by a direct model is important, the more the MSE on the shape estimated by a MC estimator is important. The IAS model is systematically less efficient than the IAF model because it has the same properties in terms of systematic un-biasedness as the IAF model, while having a larger number of unseen modes than the IAF model [13, Chap. 4].

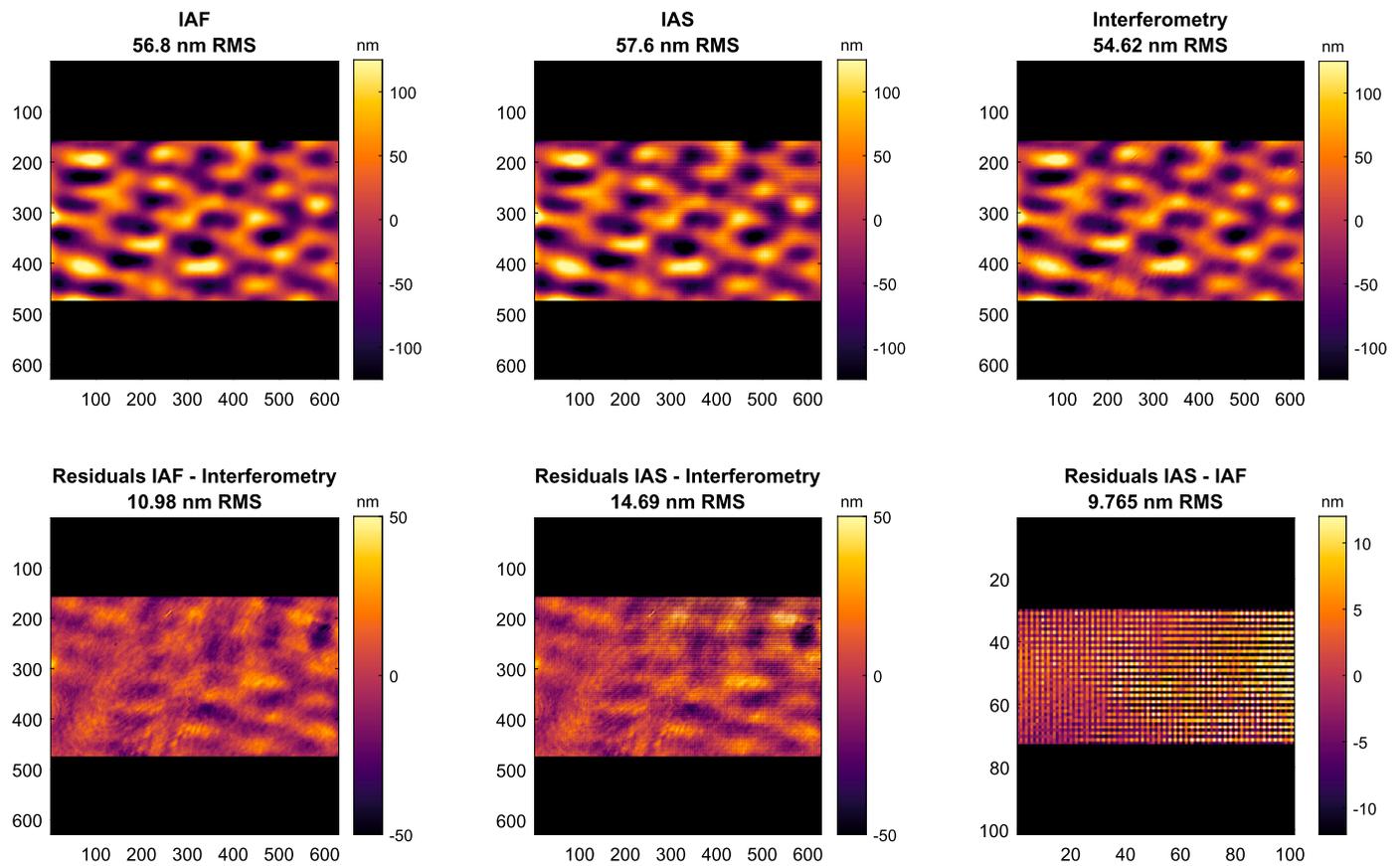

**Fig. 4.** Shape reconstruction (in nm) measured of the freeform mirror polished by SAFRAN REOSC for CNES by a MAP estimator, respectively associated with the IAF (top left), IAS (top center) models and phase shift interferometry (top right). The first 36 Zernike polynomials were subtracted from all shape maps to visualize the high frequencies of the mirror. The residuals between the shape reconstruction and interferometry are displayed in the second row (bottom left and center). The residuals between the IAS and IAF reconstruction are shown at a different scale for visualization purposes (bottom right).

The PS model paired either with the LS or MAP estimator performs worse in terms of MSE for SNR higher than 1 than the IAS/IAF models. This error is stable with different level of noise (i.e. SNR), as the systematic error due to the pre-processing of the slope data dominates the total shape reconstruction error. For a SNR of 1, because it incorporates a regularization, the PS/MAP pair performs better in terms of MSE than the PS/LC pair, but still worse than the IAS/MAP and IAF/MAP pairs. The bias term in the MSE due to the pre-processing of the slope data by the PS model makes it perform worse in terms of MSE than the unbiased models, even for an SNR of 1 and paired with a MAP estimator.

In conclusion, the use of the PS model and the IAS model is never justified and we recommend, whatever the SNR encountered experimentally is, the use of the IAF/MAP estimator couple.

## 5. Application to shape reconstruction by experimental deflectometry

We have simulated in the previous section the performance in terms of mean square error on the estimated shape of different model/estimator pairs. In order to apply our approach to experimental data, we compare in this section the shape estimated by different model/estimator pairs to data acquired during a shape measurement of a highly aspherical freeform mirror polished by SAFRAN REOSC in the framework of a research and technology project in cooperation with CNES (maximum aspherical slope greater than 50 mrad). The useful surface of the optical surface is a rectangle of 120 mm x 220 mm. The optical surface of this mirror was measured both by deflectometry and phase-shift interferometry (using a computer-generated hologram as a wavefront reference) to evaluate the quality of the MAP shape reconstruction methods. The mirror was measured during polishing, and we stress that these measurements do not represent the final quality of the polished mirror. Deflectometry setup consists of a 1280x1024 display, 19 inches Terra MF190D-L03, for which the manufacturer indicates a gamma value of 2.2. The camera used is a FLIR CM3-U3-13S2M-CS (1.3 MP, 30 FPS) with a SONY ICX445 MONO detector. The objective is a $1/1.8''$ 10-40 mm varifocal objective from Edmund Optics. Camera and display were set up next to the center of curvature of the freeform mirror (∼640 mm). The geometric setup has not been calibrated, hence positioning uncertainties lie in the order of magnitude of the mm. The experimental setup corresponds to the same setup as in Figure 3 of [22]. This setup being off-axis, the integration support of the slopes is approximately rectangular. As described in Section 3 of this paper, this rectangle integration support is included in the AIF/AIS models.

Fig. 4 shows the result of the IAF, and IAS model shape reconstructions shape reconstruction of the freeform mirror, as well as the shape obtained by phase shift interferometry. The residuals between the deflectometry-based and the phase shift interferometry measurements are shown in the second row. The residuals between the IAF and IAS reconstructions as shown on a different scale (zoomed-in) for waffle mode visualization purposes. To obtain high-frequency shape maps, all shape maps shown in Fig. 4 were filtered out of the first 36 Zernike polynomials. The PSD of the mirror shape was estimated from the phase shift interferometry measurement and is in good agreement with a power law decreasing as the square of the spatial frequency. Therefore, the regularization term has been chosen as in Equation (17).

The MAP reconstructions of IAS and IAF are in good agreement with the phase-shift interferometry data. The root mean square residuals between the IAF reconstruction and the phase shift interferometry maps are 11 nm RMS, for an initial map RMS value greater than 54 nm RMS, while the residuals between the IAS reconstruction and the phase shift interferometry maps are 14.7 nm RMS. While the residuals between the IAF reconstruction and phase-shift interferometry are dominated by registration errors and spurious bangs due to the phase-shift interferometry experimental setup, the residuals between the IAS reconstruction and phase-shift interferometry exhibit additionally a large amplitude waffle mode. The residuals between the Fried and Southwell reconstructions are dominated by this waffle mode, with a root mean square of 9.7 nm. For the IAS model, a fraction of the waffle mode is present on the shape reconstructed. We point out that this waffle mode is absent from the IAF MAP reconstruction. These results show that the MAP reconstruction method associated with the IAF model allows to obtain a total shape reconstruction error compared to phase shift interferometry lower than 11 nm RMS.

## 6. Conclusion

Shape reconstruction in deflectometry is an inverse problem, which can be decomposed into three parts: choice of a direct model *i.e.*, the data formation model, choice of a decomposition basis, and choice of an estimator. The choice of a data formation model, and of a basis for zonal methods has been considerably studied in the state of the art for the reconstruction of high spatial frequencies for deflectometry, but no study existed on the impact of the choice of a direct model on the error on the reconstructed shape. Additionally, all the shape reconstruction algorithms in deflectometry are based on the Southwell model.

In this paper, we have studied the impact of the choice of the direct model on the total shape reconstruction error. For this purpose, we have designed two original models, the Fried Integral Approach (IAF) and Southwell Integral Approach (IAS), based on an integral model in accordance with the physics of data formation. These models do not generate systematic modeling errors. This study allowed us to conclude that the Pointwise Fried (PF) model belongs to the class of IAF models and therefore does not generate any systematic error. On the contrary, the Pointwise Southwell (PS) model performs an overlooked pre-processing of the data that leads to a filtering of the data, which has led the literature, both in adaptive optics and deflectometry, to underestimate the error that this model induces on the estimation of the reconstructed shape. Moreover, the PS model does not belong to the class of integral models and generates a systematic bias, due to the hidden filtering of the data. We therefore do not recommend its use and have shown that the IAS and IAF models are to be preferred.

Concerning the third step of the inverse shape reconstruction problem, we studied the performance of several estimators from the point of view of the total shape reconstruction error, whose contributors are the systematic modeling error and the propagation of measurement noise. To stabilize the inversion and avoid the propagation of slope noise on the reconstructed shape in the shape of high spatial frequencies, we propose to use a Maximum A Posteriori (MAP) estimator based on experimentally validated assumptions on the spectral power density of the shape to be measured. By means of numerical simulations, we have demonstrated that the IAF/MAP model/estimator pair is more efficient in terms of mean square error on the estimated shape than all the other model/estimator pairs, whatever the signal-to-noise ratio considered experimentally.

We then validated the use of this IAF/MAP model/estimator pair from the point of view of the total reconstruction error on experimental deflectometry data (obtained on a freeform mirror) by comparison with another metrological mean, namely phase shift interferometry. The use of the direct Fried Integral Approach (IAF) model coupled with a MAP estimator allows us to obtain a measurement of the high spatial frequencies of the freeform mirror in agreement with phase shift interferometry.

These results open the way to more economical optical manufacturing processes of freeform mirrors, which do no rely on phase-shift interferometry and mirror-specific wavefront correctors. We underline that these models are here applied to deflectometry for optical manufac-turing, but also apply to zonal shape reconstruction for Shack-Hartmaan sensors, both for optical metrology and for adaptive optics. In particular, the IAF/MAP model/estimator pair presented in this paper has been used for deflectometry measurements within the European Extremely Large Telescope project, as a metrology tool for the manufacturing of the segments of the primary mirror [13, Chap. 6].

## CRediT authorship contribution statement

**Jonquière Hugo:** Writing – review & editing, Writing – original draft, Software, Methodology, Investigation, Formal analysis, Conceptualization. **Mugnier Laurent M.:** Writing – review & editing, Visualization, Validation, Supervision, Methodology, Formal analysis, Conceptualization. **Michau Vincent:** Investigation, Conceptualization. **Mercier-Ythier Renaud:** Supervision, Investigation, Formal analysis.

## Declaration of competing interest

The authors declare the following financial interests/personal relationships which may be considered as potential competing interests:

Hugo Jonquiere reports equipment, drugs, or supplies was provided by French Space Agency. Renaud Mercier-Ythier reports a relationship with Groupe SAFRAN that includes: employment. If there are other authors, they declare that they have no known competing financial interests or personal relationships that could have appeared to influence the work reported in this paper.

## Data availability

The authors do not have permission to share data.

## Acknowledgements

We thank CNES for its support during the polishing of the freeform mirror and for allowing us to communicate the results, as well as the anonymous reviewers for their constructive comments. We also express our gratitude to the whole SAFRAN REOSC team that helped with the measurements on the freeform mirror.

## Appendix A. IAF model equation derivation

Using the IAF geometry (Fig. 2) with a reference frame origin in $(i, j)$, the local basis function $\phi_{h,k}$ is centered in $(h, k) = (i - \frac{1}{2}, j - \frac{1}{2})$. In order to clarify the notations and to avoid using two separate frames of reference for the shape and slope sampling, the remainder of this section will move to the slope frame of reference and use half-index notations for the shape sampling. From the shape decomposition (Eq. (8)), we thus obtain:

$$s_x^{i,j} = \frac{1}{p^2} \sum_{h=\frac{1}{2}}^{N-\frac{1}{2}} \sum_{k=\frac{1}{2}}^{M-\frac{1}{2}} c_{h,k} \int_{-\frac{p}{2}}^{\frac{p}{2}} \left[ \phi_0\left(\left(i + \frac{1}{2} - h\right)p, y + (j - k)\,p\right) \right. $$
$$\left. - \phi_0\left(\left(i - \frac{1}{2} - h\right)p, y + (j - k)\,p\right) \right] dy,$$

where we choose for this modeling a shape decomposed on $(N-1)(M-1)$ points, $(N, M)$ being the number of points measuring the slopes. Due to the fact that the function $\phi_0$ has a finite support of width less than 2 pixels, a pixel of slope $(i, j)$ cannot interact with a shape function that is too distant, and we have:

$$\forall h \neq i - \frac{1}{2}, \forall k \notin \left\{j - \frac{1}{2}, j + \frac{1}{2}\right\}, \int_{-\frac{p}{2}}^{\frac{p}{2}} \phi_0\left(\left(i - \frac{1}{2} - h\right)p, y + (j - k)\,p\right) dy = 0,$$

and:

$$\forall h \neq i + \frac{1}{2}, \forall k \notin \left\{ j - \frac{1}{2}, j + \frac{1}{2} \right\}, \int_{-\frac{p}{2}}^{\frac{p}{2}} \phi_0 \left( (i + \frac{1}{2} - h)p, y + (j - k)p \right) dy = 0,$$

which leads to:

$$s_x^{i,j} = \frac{1}{p^2} \int_{-\frac{p}{2}}^{\frac{p}{2}} \left[ \sum_{k=j-\frac{1}{2}}^{j+\frac{1}{2}} c_{i+\frac{1}{2},k} \phi_0 \left( 0, y + (j - k)p \right) \right.$$

$$\left. - \sum_{k=j-\frac{1}{2}}^{j+\frac{1}{2}} c_{i-\frac{1}{2},k} \phi_0 \left( 0, y + (j - k)p \right) \right] dy,$$

or finally:

$$s_x^{i,j} = \frac{1}{p^2} \int_{-\frac{p}{2}}^{\frac{p}{2}} \left[ c_{i+\frac{1}{2},j+\frac{1}{2}} \phi_0 \left( 0, y - \frac{p}{2} \right) + c_{i+\frac{1}{2},j-\frac{1}{2}} \phi_0 \left( 0, y + \frac{p}{2} \right) - \right.$$

$$\left. c_{i-\frac{1}{2},j+\frac{1}{2}} \phi_0 \left( 0, y - \frac{p}{2} \right) - c_{i-\frac{1}{2},j-\frac{1}{2}} \phi_0 \left( 0, y + \frac{p}{2} \right) \right] dy.$$

Now, because the function $\phi_0$ is symmetrical about the $Ox$ axis, we have:

$$\int_0^p \left[ \phi_0(0, y) \right] dy = \int_{-p}^0 \left[ \phi_0(0, y) \right] dy,$$

we obtain by change of variable in the integrals and factorization:

$$s_x^{i,j} = \left( \frac{c_{i+\frac{1}{2},j+\frac{1}{2}} - c_{i-\frac{1}{2},j+\frac{1}{2}} + c_{i+\frac{1}{2},j-\frac{1}{2}} - c_{i-\frac{1}{2},j-\frac{1}{2}}}{2p} \right) \frac{2}{p} \int_0^p \left[ \phi_0(0, y) \right] dy.$$

We then define a dimensionless constant $\alpha$, as:

$$\alpha = \frac{2}{p} \int_0^p \left[ \phi_0(0, y) \right] dy.$$

And we get:

$$s_x^{i,j} = \alpha \left( \frac{c_{i+\frac{1}{2},j+\frac{1}{2}} - c_{i-\frac{1}{2},j+\frac{1}{2}} + c_{i+\frac{1}{2},j-\frac{1}{2}} - c_{i-\frac{1}{2},j-\frac{1}{2}}}{2p} \right),$$

or if one prefers conventions without half-indexes once the calculation is done, and thus taking the origin of the numbering of the sampling of the shape in $\left( -\frac{1}{2}, -\frac{1}{2} \right)$:

$$s_x^{i,j} = \alpha \left( \frac{c_{i+1,j} - c_{i,j} + c_{i+1,j+1} - c_{i,j+1}}{2p} \right), \tag{23}$$

## Appendix B. IAS model equation derivation

Using the IAS geometry (Fig. 2) and shape decomposition (Eq. (8)), we obtain:

$$s_x^{i,j} = \frac{1}{p^2} \sum_{h=1}^{N} \sum_{k=1}^{M} c_{h,k} \int_{-\frac{p}{2}}^{\frac{p}{2}} \left[ \phi_0 \left( (i + \frac{1}{2} - h)p, y + (j - k)p \right) \right.$$

$$\left. - \phi_0 \left( (i - \frac{1}{2} - h)p, y + (j - k)p \right) \right] dy.$$

Because the function $\phi_0$ has a finite support of width less than 2 pixels, a pixel of slope $(i, j)$ cannot interact with a function of shape $(h, k)$ too distant, which leads to:

$$\forall i, (i - h) \notin \{-1, 0\}, \forall j, |j - k| > 1,$$

$$\int_{-\frac{p}{2}}^{\frac{p}{2}} \phi_0 \left( (i - h + \frac{1}{2})p, y + (j - k)p \right) dy = 0,$$

and:

$$\forall i, (i - h) \notin \{0, 1\}, \forall j, |j - k| > 1,$$

$$\int_{-\frac{p}{2}}^{\frac{p}{2}} \phi_0 \left( (i - h - \frac{1}{2})p, y + (j - k)p \right) dy = 0,$$

we obtain:

$$s_x^{i,j} = \frac{1}{p^2} \sum_{h=i}^{i+1} \sum_{k=j-1}^{j+1} c_{h,k} \int_{-\frac{p}{2}}^{\frac{p}{2}} \phi_0 \left( (i - h + \frac{1}{2})p, y + (j - k)p \right)$$

$$- \frac{1}{p^2} \sum_{h=i-1}^{i} \sum_{k=j-1}^{j+1} c_{h,k} \int_{-\frac{p}{2}}^{\frac{p}{2}} \phi_0 \left( (i - h - \frac{1}{2})p, y + (j - k)p \right) dy.$$

Since the function $\phi_0$ is symmetric along $Ox$, the contributions to $s_x^{i,j}$ of the basis functions located on the $j^i$ column are zero and we have:

$$\forall k, \forall j, \int_{-\frac{p}{2}}^{\frac{p}{2}} \left[ \phi_0 \left( -\frac{p}{2}, y + (j - k)p \right) - \phi_0 \left( \frac{p}{2}, y + (j - k)p \right) \right] dy = 0,$$

and the terms in $h = i$ therefore cancel, which leads to:

$$s_x^{i,j} = \frac{1}{p^2} \sum_{j=k-1}^{k+1} c_{i+1,j} \int_{-\frac{p}{2}}^{\frac{p}{2}} \phi_0 \left( -\frac{p}{2}, y + (j - k)p \right) dy$$

$$- \frac{1}{p^2} \sum_{j=k-1}^{k+1} c_{i-1,j} \int_{-\frac{p}{2}}^{\frac{p}{2}} \phi_0 \left( \frac{p}{2}, y + (j - k)p \right) dy.$$

Fig. 5 shows the different interaction configurations as a function of the size of the $\phi_0$ support, which is called the local support. Configuration (a) presents the case where the detector pixel is inscribed in the local support. The configuration (b) presents the case where the local support is larger than the side of the detector pixel, without the detector pixel being inscribed in the local support. The case where the local support is inscribed in the detector pixel is not shown because it represents a case where the direct model is not sensitive to any local basis function. The value of the slope measured in $(i, j)$ depends on the local function $(h, k)$ if the latter's support intercepts the detector pixel (in red on Fig. 5). We thus note on Fig. 5a that in the case of a square local support and of side strictly greater than the side of the integral support, the value of the slope measured in $(i, j)$ depends on six coefficients $c_{h,k}$. On the other hand, in the configuration of Fig. 5b, the value of the slope measured in $(i, j)$ depends only on two coefficients: $c_{i-1,j}$ and $c_{i+1,j}$.

Thus, and contrary to the IAF geometrical configuration, the IAS configuration restricts the choice of basis functions to functions with a local support strictly greater than the size of the detector pixel, but without the detector pixel being inscribed in the local support if we wish to limit the internal dependencies of the direct model. We recall that the basis function $\phi_0$ must cancel at the boundary of its support to respect the continuity condition of the local basis. However, it must have a value as large as possible on the sides of the detector pixel so that the sensitivity of the IAS model to the measured shape is not negligible. This condition imposes a fast variation of the function $\phi_0$ on the interval $\left[ \frac{p}{2} : \frac{p}{\sqrt{2}} \right]$. Note that in the case of a detector pixel inscribed in the local support,

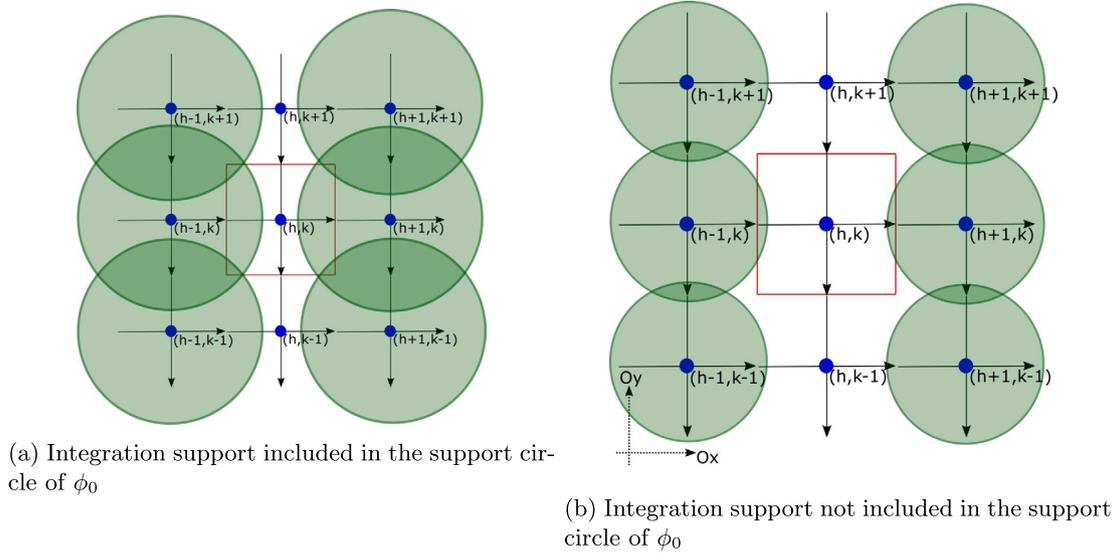

(a) Integration support included in the support circle of $\phi_0$

(b) Integration support not included in the support circle of $\phi_0$

**Fig. 5.** Dependence on local functions of the shape decomposition basis as a function of the size of the support of $\phi_0$ in the IAS model.

the contribution to the slope of the pixels in the corners is less than the contribution of the horizontally adjacent pixels, which implies that the signal-to-noise ratio of the inversion will be degraded.[1]

Subsequently, in order to avoid a model with a 6-coefficient slope dependence, we make the assumption for the IAS model of a function $\phi_0$ of local support strictly greater than the size of the detector pixel, but without the detector pixel being inscribed in the local support. It follows:

$$s_x^{i,j} = (\frac{c_{i+1,j} - c_{i-1,j}}{2p})\frac{2}{p}\int_{-\frac{p}{2}}^{\frac{p}{2}}\left[\phi_0(\frac{p}{2}, y)\right]dy.$$

We then define a constant $\beta$, which depends only on the choice of $\phi_0$, as:

$$\beta = \frac{2}{p}\int_{-\frac{p}{2}}^{\frac{p}{2}}\phi_0(\frac{p}{2}, y)dy = \frac{4}{p}\int_{0}^{\frac{p}{2}}\phi_0(\frac{p}{2}, y)dy. \quad (24)$$

Hence:

$$s_x^{i,j} = \beta(\frac{c_{i+1,j} - c_{i-1,j}}{2p}). \quad (25)$$

### Appendix C. Link between integral and point-wise models

#### C.1. Link between the PF model and the IAF model

Following the Equation (3), the derivative of $\phi$ along the direction $O_x$ evaluated at the center of the pixel is:

$$\frac{\partial\phi}{\partial x}\left(\frac{p}{2}, \frac{p}{2}\right) = a + \frac{cp}{2}.$$

And the average of this derivative over the pixel area is:

$$\frac{1}{p^2}\int_{0}^{p}\int_{0}^{p}\frac{\partial\phi}{\partial x}dxdy = a + \frac{cp}{2}.$$

Therefore, in the case of Fried's model, when using the same basis as the PF model, the IAF model estimates the same value of slopes as the PF model, and hence leads to the same discrete equation from the PF model as derived in Equation (4). The two models IAF and FP thus lead

to the same discrete equation. Moreover, if we choose $\phi_0$ as the product of two pyramidal functions:

$$\phi_0(x, y) = \frac{(p - |x|)(p - |y|)}{p^2}\Pi_p(x, y),$$

where $\Pi_p(x, y)$ is the gate function of support equal to the integration support, then the associated IAF model is equivalent to Fried's model, essentially because bi-linear splines and pyramidal functions shifted by half a pixel describe the same functional space. The Fried model can thus be interpreted as a particular integral approach model, and coincides with an IAF model for pyramidal basis functions and $\beta = 1$.

### Appendix D. Link between the PS model and the IAS model

Let us consider an IAS model with $\alpha = 1$, of discrete equation given by the equation (11):

$$s_x^{i,j} = (\frac{c_{i+1,j} - c_{i-1,j}}{2p}).$$

Then by forming an average slope in a similar way to the pre-processing of the PS model data:

$$\frac{s_x^{i,j} + s_x^{i,j+1}}{2} = (\frac{c_{i+2,j} + c_{i+1,j} - c_{i,j} - c_{i-1,j}}{2p}).$$

The IAS and PS models can only lead to the same discrete equation if Equation (1) from the discrete PS model:

$$\frac{s_x^{i,j} + s_x^{i+1,j}}{2} = (\frac{\phi_{i+1,j} - \phi_{i,j}}{p})$$

is true for all the continuous shapes to be estimated. We recall that in the framework of the AIS model, we have:

$$\phi(i, j) = c_{i,j},$$

which imposes, for the set of possible continuous shapes to be estimated:

$$\forall h, \forall k, c_{i+1,j} - c_{i,j} = c_{i+2,j} - c_{i-1,j}.$$

This constraint is not met for any general shape. Hence the IAS and PS models are different data formation models, and thus the PS model cannot be interpreted as a particular model of the IAS class.

---

[1] See section 4 of this paper.

## Appendix E. Total shape error estimated by zonal MAP reconstruction assuming periodic boundary conditions

Assuming periodic boundaries for the shape reconstruction problem, models described in Equations (11), (10), (4), and (1) can be expressed in a convolution formalism:

$$\begin{cases} s_x = h_x * \phi, \\ s_y = h_y * \phi, \end{cases} \tag{26}$$

where $h_x, h_y$ a convolution kernel specific to the model at hand. The MAP estimator can then be expressed as a Wiener filter in the Fourier domain [23, Chap. 6]:

$$\hat{\phi} = \mathfrak{F}^{-1} \left( \frac{\tilde{h_x}^* (\tilde{s_x'} + \tilde{n_x'}) + \tilde{h_y}^* (\tilde{s_y'} + \tilde{n_y'})}{|\tilde{h_x}|^2 + |\tilde{h_y}|^2 + \lambda \frac{\sigma^2}{S_\phi(k_x, k_y)}} \right),$$

where $\tilde{x} = \mathfrak{F}(x)$ the Fourier transform of a variable. The noises $n_{s,x}, n_{s,x}$ are two independent realizations of the measurement noise on the slopes and * denotes the conjugate. Replacing the slopes by their definition in Equation (26), we obtain:

$$\hat{\phi} = \mathfrak{F}^{-1} \left( \frac{\left( |\tilde{h_x}|^2 + |\tilde{h_y}|^2 \right) \tilde{\phi}}{|\tilde{h_x}|^2 + |\tilde{h_y}|^2 + \lambda \frac{\sigma^2}{S_\phi(k_x, k_y)}} \right) + \mathfrak{F}^{-1} \left( \frac{\tilde{h_x}^* \tilde{n_x'} + \tilde{h_y}^* \tilde{n_y'}}{|\tilde{h_x}|^2 + |\tilde{h_y}|^2 + \lambda \frac{\sigma^2}{S_\phi(k_x, k_y)}} \right). \tag{27}$$

We emphasize that the least-squared estimator corresponds to the case where $\lambda = 0$, and that this notation therefore covers all pairs of direct models & estimators. From Equation (27) we can calculate the mean square error $\epsilon_\phi$ on the estimated form:

$$\epsilon_\phi^2 = \frac{1}{N^2} \langle \| \hat{\phi} - \phi \|^2 \rangle, \tag{28}$$

where square brackets denote the average value over a large number of measurements. Hence:

$$\epsilon_\phi^2 = \frac{1}{N^2} \left\langle \left\| \mathfrak{F}^{-1} \left( \frac{\left( |\tilde{h_x}|^2 + |\tilde{h_y}|^2 \right) \tilde{\phi}}{|\tilde{h_x}|^2 + |\tilde{h_y}|^2 + \lambda \frac{\sigma^2}{S_\phi(k_x, k_y)}} \right) \right. \right.$$
$$\left. \left. + \mathfrak{F}^{-1} \left( \frac{\tilde{h_x}^* \tilde{n_x'} + \tilde{h_y}^* \tilde{n_y'}}{|\tilde{h_x}|^2 + |\tilde{h_y}|^2 + \lambda \frac{\sigma^2}{S_\phi(k_x, k_y)}} \right) - \phi \right\|^2 \right\rangle.$$

Assuming isotropic white noise with mean zero and standard deviation $\sigma$ on the slopes, we have according to [18, Sect 9.4]:

$$\begin{cases} \langle \tilde{n_x'} \rangle = 0, \\ \langle \tilde{n_y'} \rangle = 0, \end{cases}$$

the cross terms proportional to $\langle \tilde{n_x'} \rangle$ or $\langle \tilde{n_y'} \rangle$ cancel out and we obtain:

$$\epsilon_\phi^2 = \frac{1}{N^2} \left\langle \left\| \mathfrak{F}^{-1} \left( \frac{\left( |\tilde{h_x}|^2 + |\tilde{h_y}|^2 \right) \tilde{\phi}}{|\tilde{h_x}|^2 + |\tilde{h_y}|^2 + \lambda \frac{\sigma^2}{S_\phi(k_x, k_y)}} - \tilde{\phi} \right) \right\|^2 \right\rangle$$
$$+ \frac{1}{N^2} \left\langle \left\| \mathfrak{F}^{-1} \left( \frac{\tilde{h_x}^* \tilde{n_x'} + \tilde{h_y}^* \tilde{n_y'}}{|\tilde{h_x}|^2 + |\tilde{h_y}|^2 + \lambda \frac{\sigma^2}{S_\phi(k_x, k_y)}} \right) \right\|^2 \right\rangle,$$

which leads, using Parserval's equality [24], to:

$$\epsilon_\phi^2 = \frac{1}{N^2} \left\langle \left\| \left( \frac{\left( |\tilde{h_x}|^2 + |\tilde{h_y}|^2 \right)}{|\tilde{h_x}|^2 + |\tilde{h_y}|^2 + \lambda \frac{\sigma^2}{S_\phi(k_x, k_y)}} - 1 \right) ||\tilde{\phi}||^2 \right\rangle$$
$$+ \frac{1}{N^2} \left\langle \left\| \frac{\tilde{h_x}^* \tilde{n_x'} + \tilde{h_y}^* \tilde{n_y'}}{|\tilde{h_x}|^2 + |\tilde{h_y}|^2 + \lambda \frac{\sigma^2}{S_\phi(k_x, k_y)}} \right\|^2 \right\rangle.$$

In the case where the assumptions made about the shape to be reconstructed are correct, the shape $\phi$ of the optical part is a realization of the *a posteriori* distribution $\rho(\phi)$, and we therefore recognize the spectral power density $S_\phi$ of the shape to be reconstructed:

$$S_\phi = \langle \| \tilde{\phi} \|^2 \rangle,$$

where $\langle \cdot \rangle$ denotes the average over a large number of realizations of $\phi$, which is here a random variable following a probability distribution *a posteriori*. We identify here the two contributors to the total error on the estimated shape, the bias linked to the direct model chosen and the propagation of noise on the slope measurements to the estimated shape. The standard deviation of the noise on the estimated shape is then $sigma_\phi$ and we have:

$$\sigma_\phi^2 = \frac{1}{N^2} \left\langle \left\| \frac{\tilde{h_x}^* \tilde{n_x'} + \tilde{h_y}^* \tilde{n_y'}}{|\tilde{h_x}|^2 + |\tilde{h_y}|^2 + \lambda \frac{\sigma^2}{S_\phi(k_x, k_y)}} \right\|^2 \right\rangle,$$

using the $L - 2$ norm, we obtain:

$$\sigma_\phi^2 = \left\langle \frac{1}{N^2} \sum_{m_x = -N/2}^{N/2} \sum_{m_y = -N/2}^{N/2} \left| \left( \frac{\tilde{h_x}^* \tilde{n_{s,x}} + \tilde{h_y}^* \tilde{n_{s,y}}}{|\tilde{h_x}|^2 + |\tilde{h_y}|^2 + \lambda \frac{\sigma^2}{S_\phi(k_x, k_y)}} \right)_{k_x, k_y} \right|^2 \right\rangle,$$

where $N$ is the lateral resolution of the camera used for acquisition and where the wave numbers $(m_x, m_y)$ are related to the spatial frequencies $(k_x, k_y)$ by:

$$\begin{aligned} k_x &= \frac{2\pi m_x}{Np} \\ k_y &= \frac{2\pi m_y}{Np}. \end{aligned}$$

From [18, Sect 9.4], we obtain for a white noise statistic:

$$\langle \tilde{n}_{s,x} \times \tilde{n}_{s,y} \rangle = 0,$$

and:

$$\langle \tilde{n}_{s,x} \times \tilde{n}_{s,x} \rangle = \sigma^2.$$

Hence:

$$\sigma_\phi^2 = \frac{\sigma^2}{N^2} \sum_{m_x = -N/2}^{N/2} \sum_{m_y = -N/2}^{N/2} \left( \frac{|\tilde{h_x}|^2 + |\tilde{h_y}|^2}{\left( |\tilde{h_x}|^2 + |\tilde{h_y}|^2 + \lambda \frac{\sigma^2}{S_\phi(k_x, k_y)} \right)^2} \right)_{k_x, k_y}.$$

By replacing the norm in the bias term with its definition, we obtain the root mean square error on the shape estimated by a MAP estimator:

$$\epsilon_\phi^2 = \frac{1}{N^2} \sum_{m_x = 1}^{N} \sum_{m_y = 1}^{N} \left( \frac{\lambda \frac{\sigma^2}{S_\phi(k_x, k_y)}}{|\tilde{h_x}|^2 + |\tilde{h_y}|^2 + \lambda \frac{\sigma^2}{S_\phi(k_x, k_y)}} \right)^2_{k_x, k_y} S_\phi(k_x, k_y)$$
$$+ \frac{\sigma^2}{N^2} \sum_{m_x = 1}^{N} \sum_{m_y = 1}^{N} \left( \frac{|\tilde{h_x}|^2 + |\tilde{h_y}|^2}{\left( |\tilde{h_x}|^2 + |\tilde{h_y}|^2 + \lambda \frac{\sigma^2}{S_\phi(k_x, k_y)} \right)^2} \right)_{k_x, k_y}.$$

We emphasize again that $(\tilde{h}_x, \tilde{h}_y)$ depend on the direct model considered, and that these terms can be calculated analytically for each direct model.